# A method of laser micro-polishing for metallic surface using UV nano-second pulse and CW lasers

Pong-Ryol Jang, Tae-Sok Jang, Kum-Hyok Ji, Nam-Chol Kim

Institute of Science and Experimental Instruments, Kim Il Sung University, DPRK

**Abstract:** During laser micro-polishing, the constant control of laser energy density is a key technology to improve the surface roughness. In this paper, a method which controls the energy density of UV(ultraviolet) pulse laser in real time with the control of CW(continuous wave) laser spot size in laser micro-polishing for metallic surface was presented. The experimental and analytical considerations of several influence factors such as laser spot size, fusion zone and focal offset were investigated. In addition, using a laser micro-polishing system manufactured with this method, the laser micro-polishing experiments on the two different surface shapes of stainless steel 316L were conducted. For the inclined or curved surface, the surface roughness improvements of up to 56.4% and 57.3% were respectively obtained, and the analysis of the results were discussed.

## 1. Introduction

The laser micro-polishing has already been one important branch of research in material surface processing as a new surface processing technology. With the development of micro &nanotechnology and precision machinery industry, the polishing technology is becoming growing in demand as time go on, producing products of high quality produced, reducing product costs and having multi functions.

A lot of researchers are working on various kinds of materials, especially on the laser polishing mechanism of different metals and the choice of the laser polishing methods and the analysis of factors that have been known to affect the polishing effects with laser polishing technique. The factors such as laser energy density, wavelength, pulse duration, angle of incidence, scanning speed, scanning method, and characters of tested material significantly affect the features of laser polishing process [1-5]. Studies show that proper adjustment of laser processing parameters and moving the stage, can reduced the surface roughness [6,7].

It is one of the most important to choose the optimum laser energy density in laser polishing process. The focal offset measured from the focal position of objective lens to the part surface affects directly to the laser spot size on the material surface, is one of the most difficult parameters to control in laser micro-polishing. 3D surfaces are polished by using a system that is similar to applied in industrial five-axis laser cutting systems and the trajectories are programmed by a CAM system [8].

For the micro-polishing on the metallic surface such as stainless steel, this paper describes a method for real time control of laser energy density on the workpiece using two lasers. And the experimental and analytical considerations of several factors for the method were investigated. Using the new method the laser micro-polishing experiments on the inclined and curved surface of stainless steel 316L were conducted, and the analysis of the results were discussed.

## 2. Micro-polishing by two lasers

### 2.1. Method

The difference between laser micro-polishing and other laser process is related to the comparatively large laser spot on the workpiece surface and strict demand for the constant laser energy density. Especially in laser micro-polishing on incline or curved surfaces, the laser energy density shoud be controlled in real time. In the laser micro-polishing process, the laser energy directly affects effectiveness of the polishing, and assuming constant the incident laser energy, the laser energy density on the workpiece surface can be characterized by the laser spot size. Therefore the real-time control method of laser spot size on the surface is able to improve the effectiveness and efficiency of laser polishing.

In the image acquisition system that is only required the gray image, it should adopt the binarization image processing to improve processing speed and reduce costs. Using the binarization method of CCD video signal and FPGA processing technology [9], it's able to detect and control the laser spot size on the workpiece surface in real time to achieve the purpose of constant laser energy density.

During laser micro-polishing, in order to let control laser spot size by the CCD-FPGA control system, need to use two lasers. Because the high power UV pulse laser is generally used in laser micro-polishing and the repetition rate of the laser pulse is generally from tens to thousands Hz. So the frequency components of the laser pulse may be contained in CCD video signal. It is very difficult to image and control the UV laser spot on the workpiece surface by using CCD in real time. The UV pulse laser energy density is smaller than that of other laser processing and then doesn't generate plasma in the laser micro-polishing. If the laser for the spot control is the continuous visible light, then it is possible to detect the laser spot by using the CCD-FPGA control system in real time.

Fig. 1 illustrates the fundamental principle of the laser micro-polishing using two lasers.

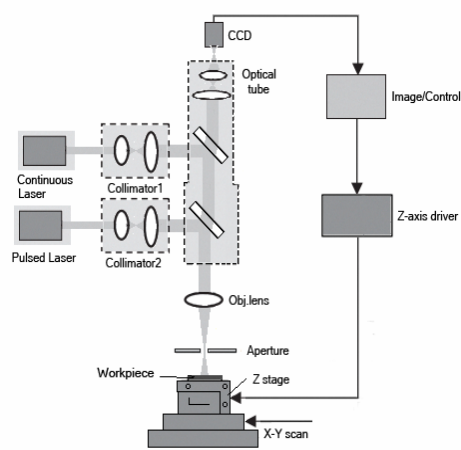

Fig. 1  The schematic of laser micro-polishing system using two lasers

The beam diameters of the CW laser and the UV pulse laser throughout the collimator are the same, the beams are defocused on the workpiece surface through the same optical system. During the two laser beams through the objective lens, their path would difference for the refractive index difference because of the wavelength of the two lasers, but it can be compensated by the color compensation objective lens. Therefore, the two laser spot sizes which are set on the workpiece surface are the same. The CW laser spot reflected on the surface returns is imaged in CCD also through the optical system to be processed in the Image/Control unit which FPGA is the core of. The output signals of the FPGA drive the Z axis stage, consequently the feedback control is achieved. Therefore the low-power CW laser can form the stable laser spot on the workpiece surface, by accuracy measurement and control of the laser spot size, the purpose for constant energy density of the high power UV pulse laser irradiated on the workpiece surface is achieved.

**2.2. The spot size of CW laser and the fusion zone size by the UV pulse laser**

During laser micro-polishing, the UV pulse laser energy density on the workpiece surface is smaller than of other laser processing not to produced plasma. So the UV pulse laser does not affect on the image of CW laser spot on the workpiece surface. In the structure of the optical system, the diameters of the two laser spots are equal on the workpiece surface. But actually the fusion zone size by the UV pulse laser and the spot size of the CW laser are different during the laser micro-polishing with the two lasers.

Fig. 2 shows the energy distribution of the UV pulse laser on the workpiece surface. When the mode of laser is $TEM_{00}$, its energy distribution is a Gaussian profile(Eq. (1)), is written as following equation [10].

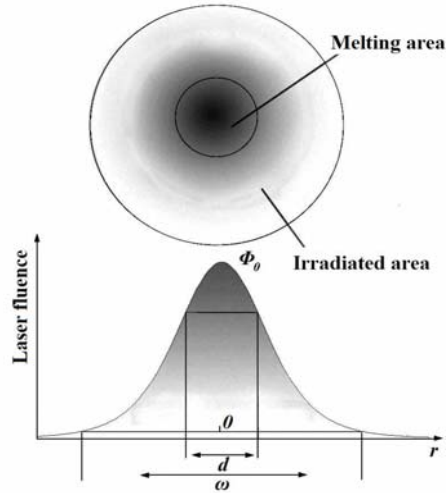

Fig. 2   The energy density distribution of UV pulse laser

$$\phi(r) = \phi_0 \exp(-\frac{8r^2}{\omega^2}) \quad (1)$$

where $\omega$ is laser spot diameter, $r$ is distance from the spot center, and $\phi_0$ is maximum energy density of the laser beam which is irradiated on the surface. The relationship between $\phi_0$ and the pulse laser energy $E_0$ is written as [11]:

$$\phi_0 = \frac{8E_0}{\pi \times \omega^2} \quad . \quad (2)$$

From Eq. (1), the pulse laser energy density threshold for laser micro polishing is written as:

$$\phi_{th} = \phi_0 \exp(-\frac{8r^2}{\omega^2}), \quad (3)$$

where $\phi_{th}$ is pulse laser energy density threshold.

Therefore, when the diameter of the fusion zone is *d*, it is expressed as

$$d^2 = 2\omega^2 \ln\frac{\phi_0}{\phi_{th}}. \qquad (4)$$

Hence the fusion zone size is given as:

$$s = 2s_0 \ln\frac{\phi_0}{\phi_{th}}, \qquad (5)$$

where *s* is size of fusion zone, $s_0$ is CW laser spot size( it is equal to the UV laser spot's ).

As can be known in Eq. (5), during the UV pulse laser energy density is controlled in real time and the energy density threshold is constant for the material in the laser micro-polishing, the relationship between the fusion zone size and the CW laser spot size is a linear proportion.

## 2.3. Focal offset selection

After determining the laser energy density threshold for the tested material, it is very important to select the right laser spot size. In general, the focal offset of the laser beam determines the laser spot size on the surface and cause a significant effect on results of the laser micro-polishing.

The focal offset can be situated in three distinct positions. It is considered as zero when it is set on the workpiece surface and above or below the surface was considered as positive and negative focal position, respectively(see Fig. 3). The laser energy density gradient toward the optical axis, for the case a) and b), are opposite each others.

During laser micro-polishing, especially in case of metallic surface, SSM (Surface Shallow Melting) mechanism has particular deployment [2], the case of a) is generally used. In the case of b), the gradient direction opposites against with the normal of the surface. So when the asperities of surface are melted, the concavities are simultaneously melted or evaporated, the surface roughness is consequently increased.

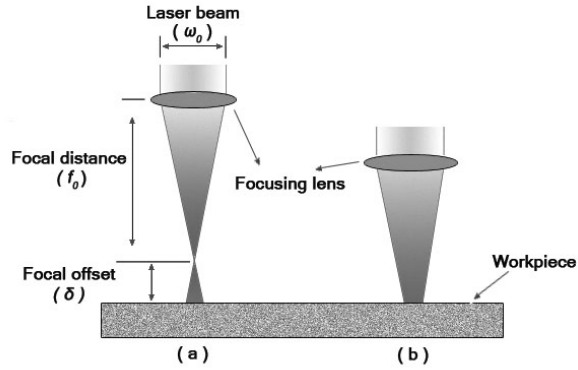

Fig. 3   Two methods to set defocus in laser polishing

When the focal offset distance is much longer than the depth of focus of the objective lens and the color compensation objective lens is used, the laser spot diameter on the workpiece surface is determined with the focal offset as:

$$\omega = \frac{\omega_0}{f_0} \times \delta, \qquad (6)$$

where $\omega$ is pulse laser spot diameter, $\delta$ is focal offset, $\omega_0$ is diameter of pulse laser beam before focused, and $f_0$ is the focal distance of the objective lens. From Eq. (1), Eq. (2) and Eq. (6), we can rewrite the pulse laser energy density as:

$$\phi(r,\delta) = \frac{8f_0^2 E_0}{\pi \omega_0^2 \delta^2} \exp(-\frac{8f_0^2 r^2}{\omega_0^2 \delta^2}). \qquad (7)$$

Fig. 4 shows the schematic of the asperities irradiated by the UV pulse laser, where $h$ is height of micro-asperity, $\delta_0$ is distance from the focus position to the workpiece surface, and $\delta_{th}$ is distance from the focus position to the fusion position of asperity.

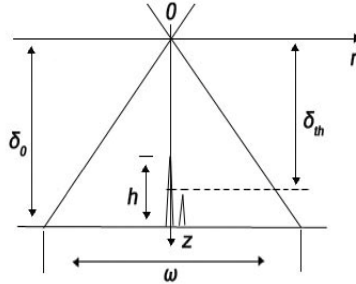

Fig. 4 schematic of the asperities irradiated by pulse laser

In $r = 0$, the laser energy density threshold ($\phi_{th}$) which the top of the micro asperity begins to be melted is written as:

$$\phi_{th} = \phi(r=0;\delta) = \frac{8f_0^2 E_0}{\pi \omega_0^2 \delta_{th}^2}. \qquad (8)$$

From Eq. (8), the distance from the focus to the fusion position of asperity can be rewritten as:

$$\delta_{th} = \sqrt{\frac{8}{\pi}} \frac{f_0}{\omega_0} \sqrt{\frac{E_0}{\phi_{th}}}. \qquad (9)$$

From the analysis of the Eq. (9), the following conclusions can be educed in the connection with laser micro-polishing.

The $\delta_0$ should be chosen as $\delta_0 > \delta_{th}$ for obtaining the good polishing effect during laser micro-polishing. When $\delta_0 - h < \delta_{th}$, the parts of the micro asperities of $h > \delta_0 - \delta_{th}$ are melted, the melted height $H$ of the micro asperities can be expressed as:

$$H = h - (\delta_0 - \delta_{th}). \qquad (10)$$

When $\delta_0 - h > \delta_{th}$, the micro asperities would not be melted.

From the above considerations, we can see that the highest height of asperities is $\delta_0 - \delta_{th}$ after the laser micro-polishing, the smaller the difference the lower the surface roughness. And in order to make the surface roughness is less than $a$, the focal offset should be precisely controlled to $\delta_0 \approx \delta_{th} + a$.

## 2.4. Objective lens and pulse laser energy density

In order to control the laser energy density in real time during the laser micro-polishing, it is very important to select the optimum objective lens. The energy density distribution of laser beam throughout the objective lens of which the focal length is different each other is given by Eq. (7).

Fig. 5 a) and b) shows the distributions energy density of laser beam for the different objective lens in the same point from the workpiece surface. The distributions energy density of laser beam throughout the objective lens of which the magnification is larger is more uniform than of the low-magnification lens in the same defocus point ( in Fig. 5 a) and b) ). However, when the laser spot sizes are same on the workpiece surface after the laser beams through the two lens each other, the distribution energy density are equal( in Fig. 5 c) and d) ).

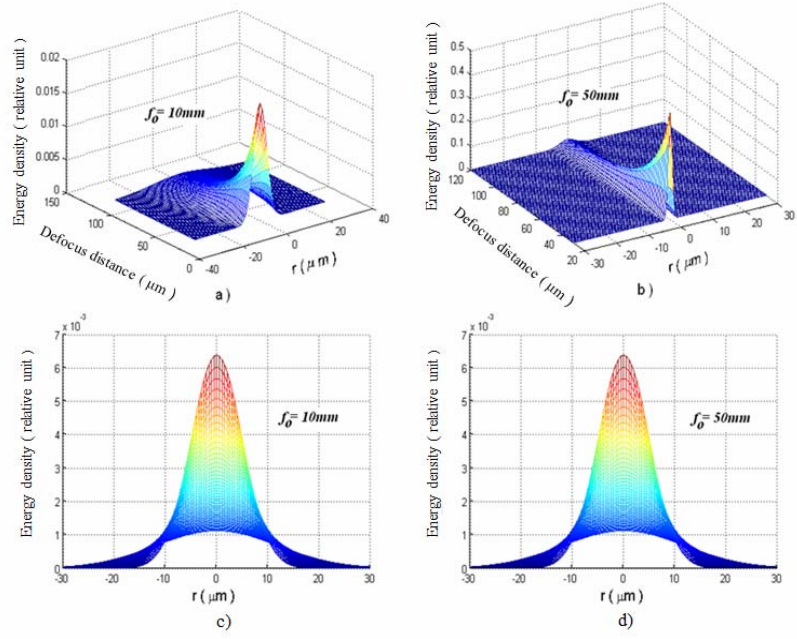

Fig. 5 Distributions of laser energy density for the different objective lens
a) and b)  For the different objective lens in the same point from the workpiece surface
c) and d)  When the spots size by the different lens are same

The vertical gradient of laser energy density after the laser beam through the objective lens is given as follows:

$$\frac{\partial \phi(r,\delta)}{\partial \delta} = \frac{2KE_0}{\pi \delta^3}(-1+\frac{Kr^2}{\pi \delta^2})\exp(-\frac{Kr^2}{\delta^2})  \qquad (11)$$

whereas $K=8\times (f_0/D_0)^2$.

When the laser spots size by the different lens are same on the workpiece surface, the gradient of the high-magnification lens is 4.5 times larger than of the low-magnification lens (Fig. 6).

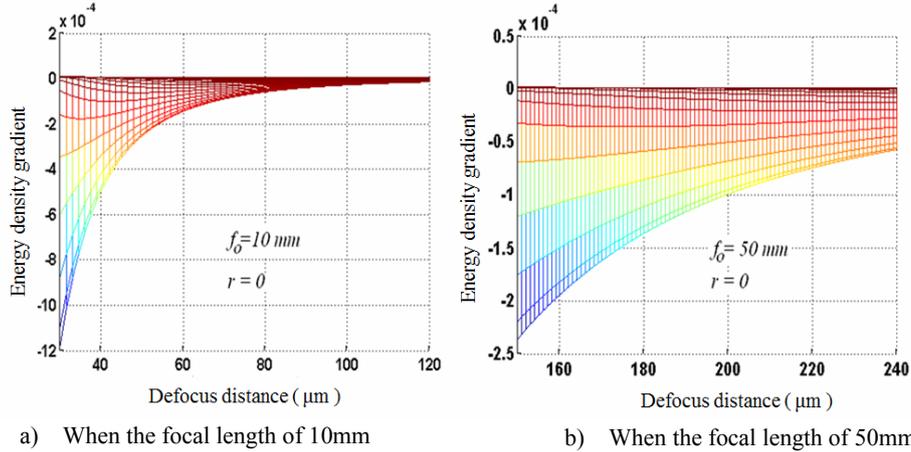

a)  When the focal length of 10mm           b)  When the focal length of 50mm

Fig. 6 The vertical gradient of laser energy density on the surface for the different focal objective lens

Therefore, in order to control accurately the laser energy density in laser micro-polishing, it should select the high-magnification lens.

## 3. Experiment and result

The used CW laser is the semiconductor laser which wavelength of 532nm, power of 10mW, beam diameter of 2.0mm and $TEM_{00}$. The UV pulse laser is the high-power Q-Switched Ultraviolet Laser (AVIA355-3000) which wavelength of 355nm, average power of 3.0W, pulse width of 40ns, beam diameter of 3.0mm and $TEM_{00}$. The optical system is the Navitar Zoom 6000's total focus optical system( the objective lens: focus of 10mm, multiple of 20, depth of focus 1.75μm and color compensation for 355nm wavelength), the collimators for the BXZ-355-1-3X and BXZ-532-1-3X (adjustable beam diameter). The CCD is the Polytec's VCT 72, the Z axis stage is the SIGMA's SGSP40-5ZF ( 0.5μm/step, 2mm/s and 5-phase stepper motor).

The workpieces are the two different surface shapes of stainless steel 316L, which one of them, sample 1, is a planar shape and the other, sample 2, is the inclined and curved surface shape which inclined angle of 8°, curved radius of less 5mm and thickness of 1mm. These were polished by sandpaper in water, coarsely polished by flatting varnish and dried after washing with ethanol.

### 3.1. Focal offset effects

In practical applications, it is very difficult to control the focal offset in real time. If the focal offset is too long, the laser spot size would be becoming further large, the laser energy density on the surface is not uniformly distributed because the energy density distribution in laser spot is a Gaussian profile, sequentially the polishing effect is degraded. And when the focal offset is too short, not only the control accuracy of laser energy density would be decreased, but also the period of polishing process is becoming longer, the polishing effect is identically degraded.

The relationship between the diameter *d* of fusion zone on the workpiece surface and the focal offset *δ* can be obtained using Eq. (5) and (6).

$$d = \frac{\sqrt{2}\omega_0}{f_0} \delta \sqrt{\ln(\frac{\phi_0}{\phi_{th}} \times \frac{f_0^2}{\delta^2})} \qquad (12)$$

The change of the diameter of fusion zone with the varying focal offset was shown in Fig. 7. Where the irradiated pulse laser energy is 150μJ, the diameter of laser beam is 4mm, the pulse number is 1, and the sample 1 was tested, the size of fusion zone is measured with 3D digital microscopy ( KH-7700, 7000 times ).

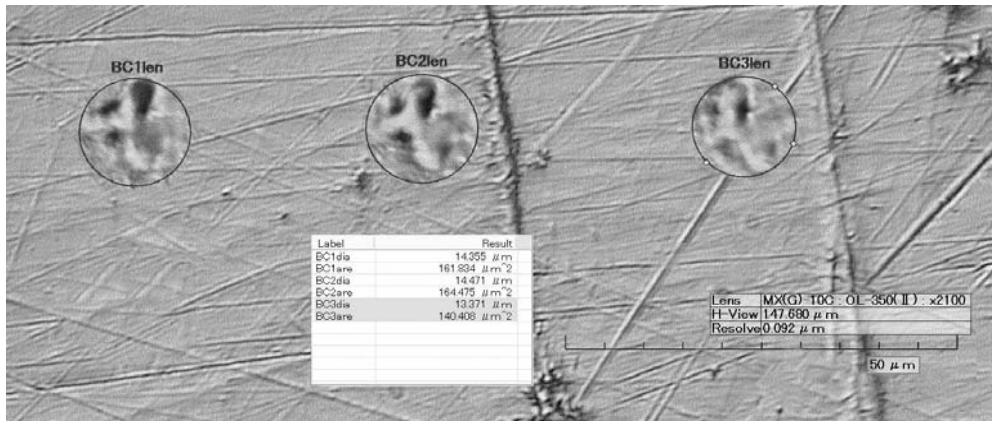

Fig. 7 The change of fusion zone with focal offset

In Fig. 8，the diameter of fusion zone is plotted for focal offset . As shown in Fig. 8 a), when the pulse energy rated, there is the optimum focal offset that the fusion zone is the largest. And as shown in Fig. 8 b), the greater the pulse energy, the larger the fusion zone of the optimum focal offset.

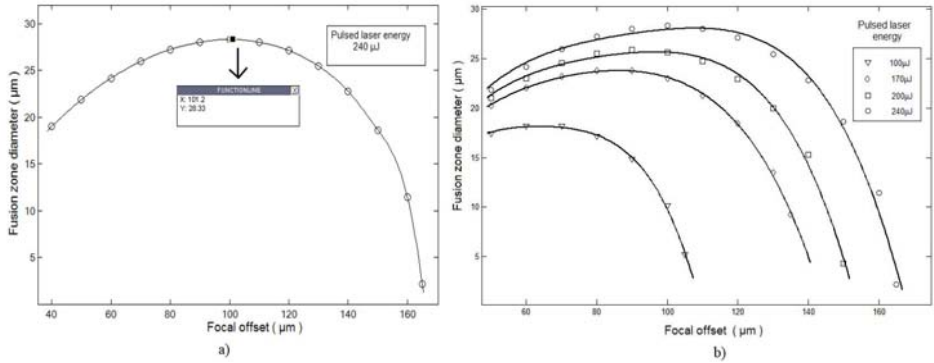

Fig. 8 The change of fusion zone diameter with the change in focal offset
a) for pulse laser energy 240μJ , b) for different pulse laser energies

Therefore the every laser energy has its corresponding optimum focal offset. The pulse laser energies and the corresponding optimum focal offset and maximum fusion zone diameter were shown in Table. 1.

Table. 1   The measured optimum data

| Pulse laser energy ( μJ ) | Optimum focal offset ( μm ) | Maximum fusion zone diameter ( μm ) |
| --- | --- | --- |
| 100 | 63.8 | 18.1 |
| 170 | 86.1 | 23.7 |
| 200 | 96.7 | 25.6 |
| 240 | 107.6 | 28.1 |

On the base of the data of Table 1, the laser micro polishing experiments had been carried out for sample 1, the results were plotted in Fig. 9.

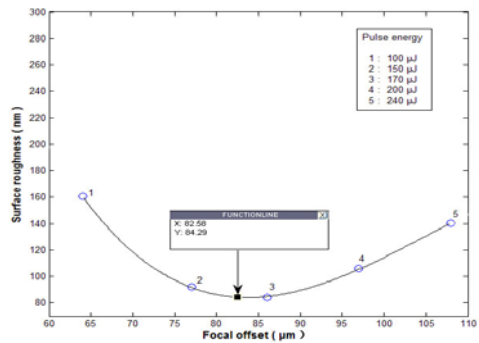

Fig. 9   The change of surface roughness with the changes in focal offset and its corresponding

As shown in Fig. 9, for given UV pulse laser energy density, there is the optimum focal offset which the best polishing result is obtained. The smallest roughness (84.29nm) was obtained in focal offset of 82.58μm in this experiment.

### 3.2. Laser micro-polishing on inclined and curved surfaces

It is the most important to ensure a constant laser energy density on the workpiece surface during laser micro-polishing of the inclined or curved surface. Because the laser spot size on the surface would be changed with the shape of surface during laser micro-polishing

The shape of the laser spot on the workpiece surface will be related with the slopes and the curvature of the surface during the laser micro-polishing on the inclined or curved surface.

The shape of laser spot is a round in the planar surface polishing, but may be a elliptic or non-symmetric elliptic in the inclined or curved surface polishing. In the case of constant laser energy (before through the objective lens), when the spot size is constant by controlling of the focal offset, the energy density on the surface is only related with the laser spot size, not be related with the shape of surface.

The scanning mode of laser beam is also important in the laser micro-polishing on the inclined or curved surface. If the slope gradient toward the scanning direction is large, then the rate of change in spot size would be increase during the process of scanning, to decrease the control accuracy of the laser energy density. Therefore the scanning direction must be set toward the inclined gradient is minimum.

Fig. 10 illustrates the laser beam scanning strategy.

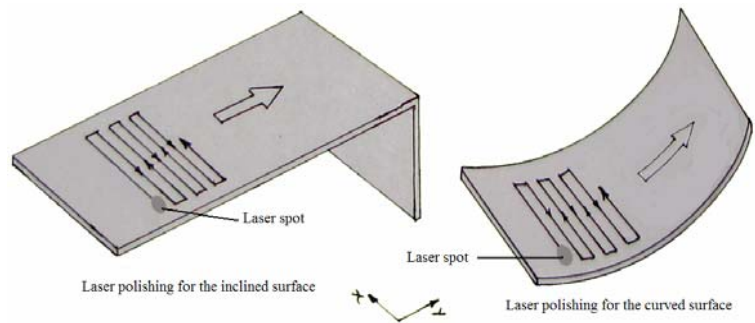

Fig. 10　Scanning methods for the inclined and curved surfaces

The experimental parameters are shown in Table. 2.

Table. 2　The experimental parameters

| Test condition | Value |
| --- | --- |
| Laser energy density | 0.14J/cm$^2$ |
| focal offset | 82μm |
| scanning speed | 18.6mm/min |
| laser spot overlap | 78% |
| pulse rate | 20Hz |

During real-time control of laser energy density, it used the reference spot size of 965μm$^2$, laser beam diameter of 4mm.

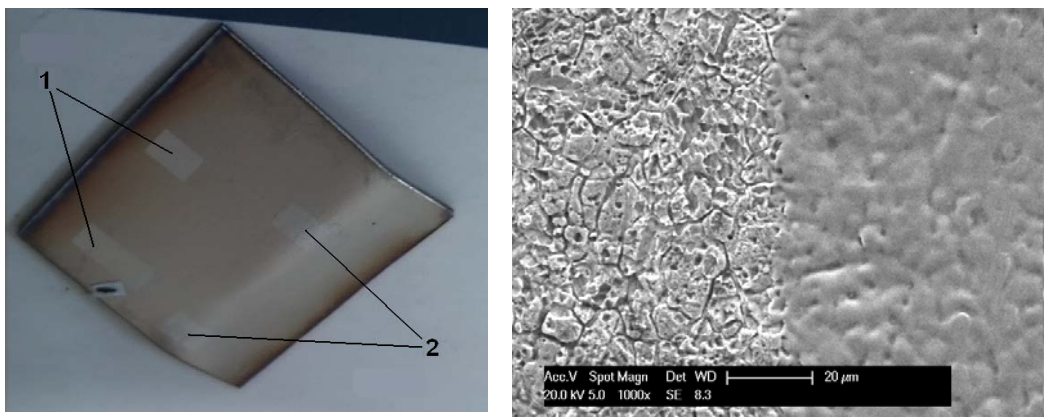

　　　　a) 3D test sample　　　　　　　　　　　　　　　b) image with SEM

Fig. 11　Results of laser micro-polishing on the sample 2

The results of laser micro-polishing on sample 2 measured using the SEM (XL-30, FEI corp.) is shown in Fig. 11. In Fig. 11 a), where '1' for the micro-polished inclined surface, '2' for the micro-polished curved surface, respectively. In Fig. 11 b), the half part at the right is polished. The roughness measurements show the roughness values below 85.70 nm Ra for the inclined surface, which means a 43.6% of the initial surface roughness, i.e. 56.4% reduction, and below 82.30 nm Ra, 57.3% reduction for the curved surface.

The results of experiment show that the slopes and curvature of the surface affect to control of laser energy density during laser micro-polishing. If the slopes or curvature too large, the accuracy of the measured spot size would be decreased so that the error of real-time control of the energy density would be increased, even cause the control mistakes.

The scanning speed of laser spot on the inclined or curved surface is faster than of the planar surface and the overlap of the Y direction also larger. Because the shape of spot on the surface would be changed, i.e. the Y direction diameter of the spot becomes smaller than the X direction diameter.

## 4. Conclusions

In this paper, a technique for stability of laser energy density in the laser micro-polishing using UV pulse laser on the material surface is described. The technique utilizes two lasers, a UV pulse laser for polishing and a CW laser for control, to control the laser spot size on the material surface by detection and control of focal offset, resulting in the constant UV pulse laser energy density on surface and more improved the laser micro-polishing efficiency.

The experimental and analytical considerations of several factors for the method were investigated. When the pulse energy rated in laser micro-polishing, there is the focal offset that the fusion zone is the largest, every pulse laser energy has its corresponding optimum focal offset. And for given pulse laser energy density and material surface, there is the focal offset in which the best polishing result is obtained.

The laser micro-polishing experiments on the two different surface shapes of stainless steel 316L were conducted. The roughness measurements show the roughness reduction of 56.4% for the inclined surface and of 57.3% for the curved surface.